\documentclass[pre,twocolumn,showpacs,showkeys,superscriptaddress,preprintnumbers,floatfix]{revtex4-1}

\usepackage{etex}
\usepackage{ifpdf}
\usepackage{hyperref}
\usepackage{dcolumn}
\usepackage{url}
\usepackage{amsmath}
\usepackage{amscd}
\usepackage{amsfonts}
\usepackage{amssymb}
\usepackage{amsthm}
\usepackage{xfrac}
\usepackage{braket}
\usepackage{bm}   
\usepackage{bbm}
\usepackage{verbatim}
\usepackage{stmaryrd}
\usepackage{xcolor}
\usepackage{setspace}
\usepackage{tikz}
\usetikzlibrary{arrows}
\usetikzlibrary{automata}
\usetikzlibrary{calc}
\usetikzlibrary{decorations.pathreplacing}
\usetikzlibrary{patterns}
\usetikzlibrary{positioning}
\usetikzlibrary{plotmarks}
\usetikzlibrary{shapes}

\usepackage{pgfplots}
\pgfplotsset{compat=newest}

\tikzstyle{vaucanson}=[
  node distance=2.5cm,
  bend angle=15,
  auto,
  every loop/.style={},
  every edge/.style={->,draw=black,line width=1.2,>=latex,shorten <=1pt,shorten >=1pt},
  every state/.style={draw=black,line width=2,font=\large},
  loop right/.style={right,out=22,in=-22,loop},
  loop above/.style={above,out=112,in=68,loop},
  loop left/.style={left,out=202,in=158,loop},
  loop below/.style={below,out=292,in=248,loop},
  loop above right/.style={right,out=67,in=23,loop},
  loop above left/.style={left,out=157,in=113,loop},
  loop below left/.style={left,out=247,in=203,loop},
  loop below right/.style={right,out=337,in=293,loop},
]

\theoremstyle{plain}    
\theoremstyle{plain}    
\theoremstyle{plain}    
\theoremstyle{plain}    
\theoremstyle{plain}    
\theoremstyle{plain}    
\theoremstyle{plain}    
\theoremstyle{plain}    
\theoremstyle{plain}    
\theoremstyle{plain}    
\theoremstyle{plain}    
\theoremstyle{plain}    
\theoremstyle{plain}    
\theoremstyle{plain}    
\theoremstyle{plain}    
\theoremstyle{plain}    
\theoremstyle{plain}

\newcommand{\eM}     {\mbox{$\epsilon$-machine}}

\newcommand{\LEN}{\ell}

\newcommand{\MeasAlphabet}  {\mathcal{A}}
\newcommand{\MeasSymbol}   { {X} }
\newcommand{\meassymbol}   { {x} }

\newcommand{\Past} { \smash{\overleftarrow {\MeasSymbol}} }
\newcommand{\past} { \smash{\overleftarrow {\meassymbol}} }

\newcommand{\Future}   { \smash{\overrightarrow{\MeasSymbol}} }
\newcommand{\future}   { \smash{\overrightarrow{\meassymbol}} }

\newcommand{\PastL} { \smash{\overleftarrow {\MeasSymbol}{}^\LEN} }

\newcommand{\FutureL}     { \smash{\overrightarrow{\MeasSymbol}{}^\LEN} }

\newcommand{\CausalState}   { \mathcal{S} }

\newcommand{\causalstate}   { \sigma }
\newcommand{\CausalStateSet}    { \boldsymbol{\CausalState} }
\newcommand{\AlternateState}    { \mathcal{R} }

\newcommand{\AlternateStateSet} { \boldsymbol{\AlternateState} }

\newcommand{\Prob}      {\Pr} 

\newcommand{\Cmu}       {C_\mu}

\newcommand{\EE}        {{\bf E}}

\newcommand{\ProcessAlphabet}   {\MeasAlphabet}

\newcommand{\forward}{+}
\newcommand{\reverse}{-}
\newcommand{\forwardreverse}{\pm} 

\newcommand{\FutureCausalState} { {\CausalState}^{\forward} }

\newcommand{\PastCausalState}   { {\CausalState}^{\reverse} }

\newcommand{\lastindex}[2]{
  \edef\tempa{0}
  \edef\tempb{#2}
  \ifx\tempa\tempb
    \edef\tempc{#1}
  \else
    \edef\tempa{0}
    \edef\tempb{#1}
    \ifx\tempa\tempb
      \edef\tempc{#2}
    \else
      \edef\tempc{#1+#2}
    \fi
  \fi
  \tempc
}

\newcommand{\I}{\mathbf{I}}

\newcommand{\CSjoint}[1][,]{
   \edef\tempa{:}
   \edef\tempb{#1}
   \ifx\tempa\tempb
      \ensuremath{\FutureCausalState\!#1\PastCausalState}
   \else
      \ensuremath{\FutureCausalState#1\PastCausalState}
   \fi
}

\newif\ifpm
\edef\tempa{\forwardreverse}
\edef\tempb{\pm}
\ifx\tempa\tempb
   \pmfalse
\else
   \pmtrue
\fi

\newcommand{\MISI}{ \overline{T} }

\renewcommand{\H}{\operatorname{H}}
\renewcommand{\I}{\operatorname{I}}

\colorlet {R_color}    {blue}
\colorlet {k_color}    {black!30!green}

\def\clap#1{\hbox to 0pt{\hss#1\hss}}

\begin{document}

\title{Statistical Signatures of Structural Organization:\\
The case of long memory in renewal processes}

\author{Sarah E. Marzen}
\email{smarzen@berkeley.edu}
\affiliation{Department of Physics\\
University of California at Berkeley,
Berkeley, CA 94720-5800}

\author{James P. Crutchfield}
\email{chaos@ucdavis.edu}
\affiliation{Complexity Sciences Center and Department of Physics\\
University of California at Davis, One Shields Avenue, Davis, CA 95616}

\date{\today}
\bibliographystyle{unsrt}

\begin{abstract}
Identifying and quantifying memory are often critical steps in developing a
mechanistic understanding of stochastic processes. These are particularly
challenging and necessary when exploring processes that exhibit long-range
correlations. The most common signatures employed rely on second-order temporal
statistics and lead, for example, to identifying long memory in processes with
power-law autocorrelation function and Hurst exponent greater than $1/2$.
However, most stochastic processes hide their memory in higher-order temporal
correlations. Information measures---specifically, divergences in the mutual
information between a process' past and future (excess entropy) and minimal
predictive memory stored in a process' causal states (statistical
complexity)---provide a different way to identify long memory in processes with
higher-order temporal correlations. However, there are no ergodic stationary
processes with infinite excess entropy for which information measures have been
compared to autocorrelation functions and Hurst exponents. Here, we show that
fractal renewal processes---those with interevent distribution tails $\propto
t^{-\alpha}$---exhibit long memory via a phase transition at $\alpha = 1$.
Excess entropy diverges only there and statistical complexity diverges there
and for all $\alpha < 1$. When these processes do have power-law
autocorrelation function and Hurst exponent greater than $1/2$, they do not have
divergent excess entropy. This analysis breaks the intuitive association
between these different quantifications of memory. We hope that the methods
used here, based on causal states, provide some guide as to how to construct
and analyze other long memory processes.
\end{abstract}


\preprint{Santa Fe Institute Working Paper 15-12-XXX}
\preprint{arxiv.org:1512.XXXX [physics.gen-ph]}

\maketitle


\setstretch{1.1}

\newcommand{\Abet}{\ProcessAlphabet}
\newcommand{\MS}{\MeasSymbol}
\newcommand{\ms}{\meassymbol}
\newcommand{\SSet}{\CausalStateSet}
\newcommand{\St}{\CausalState}
\newcommand{\st}{\causalstate}
\newcommand{\MxSt}{\AlternateState}
\newcommand{\MxSSet}{\AlternateStateSet}
\newcommand{\mxst}{\mu}
\newcommand{\mxstt}[1]{\mu_{#1}}
\newcommand{\StartMS}{\bra{\delta_\pi}}
\newcommand{\Ipred}{\EE}
\newcommand{\ISI} { \xi }

\newcommand{\ECT}{\widehat{\EE}}
\newcommand{\CCT}{\widehat{C}_\mu}


\vspace{0.2in}
\section{Introduction}

Many time series of interest have ``short memory'', meaning (loosely speaking)
that knowledge of the past confers exponentially diminishing returns for
predicting the future. However, many other time series of interest---those with
``long memory''---exhibit intrinsic timescales that grow without bound as the
amount of available data increases
\cite{Shim92a,Binn92a,Alve05a,Voss75a,dAmi86a,Pres78a}. Examples include the
hydrological data first studied by Hurst \cite{Hurs51a} and modeled by
Mandelbrot \cite{mandelbrot1968noah} and many others, e.g., see Refs.
\cite{daley1999hurst,Bera13a}.

These are qualitatively different processes that demand qualitatively different
generative models. In other words, signatures of long memory imply a kind of
structural organization of the underlying process that differs from one with
short memory. This is the \emph{inverse problem} of long memory: Which
statistical signatures identify, uniquely or not, which intrinsic
organizations? Sharp answers are critical to successful empirical analysis and
often provide necessary first steps in predictive theory building. The
complementary \emph{forward problem}, an open question, is to identify the
kinds of memoryful process structure that lead to one or another statistical
signature. Answering this question requires defining statistical signatures
that quantify memory in stochastic processes.

Many existing quantifications of long memory are based on second-order
statistics; e.g., on using the autocorrelation function, power spectrum, or
Hurst exponent. These approaches have had notable successes in analyzing
hydrological data \cite{Hurs51a,daley1999hurst}, music \cite{Voss75a}, spin
systems \cite{Binn92a}, astrophysical flicker noise \cite{Pres78a}, language
\cite{Zipf35a,Mand53a}, natural scenery \cite{Knil90a,Kuma93c}, communication
system error clustering \cite{Mand65a}, financial time series, and many other
seemingly complex phenomena \cite{dAmi86a,Mand99a}.

\begin{figure}
\centering
\includegraphics[width=0.2\textwidth]{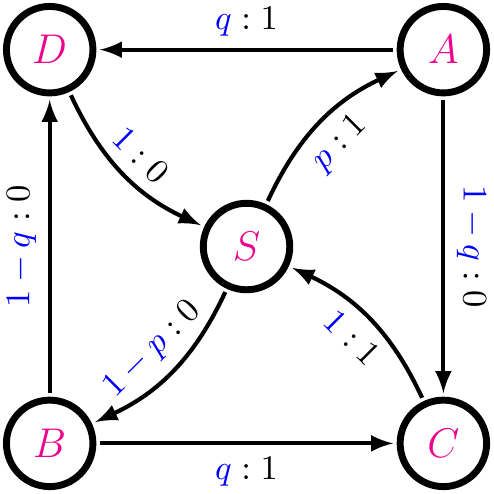}
\caption{The Random-Random-XOR (RRXOR) Process is generated by the five-state
  (minimal unifilar) Hidden Markov model here. Labels $p|\ms$ denote that a
  state-to-state transition occurs with probability $p$ and emits symbol
  $\ms$. If $X_t$ is the random variable at time $t$, then the generated time
  series is $X_{t+2} = X_{t+1} ~\text{XOR}~ X_{t}$, with $X_{t+1}$ and $X_t$
  being Bernoulli(q) and Bernoulli(p), respectively, for $t = 0, 3, 6, \ldots$.
  With $p = q = 1/2$ and starting state probabilities $\Pr(S) = 1/3$ and
  $\Pr(A) = \Pr(B) = \Pr(C) = \Pr(D) = 1/6$ the output process is stationary
  white noise---a flat power spectrum \cite{Riec15a}.
  }
\label{fig:RRXOR}
\end{figure}

However, there are at least two reasons to look to other statistics besides the
Hurst exponent.  First, second-order statistics alone can be misleading, as a
process can ``hide'' signatures of long memory in higher-order statistics. For
example, Fig.~\ref{fig:RRXOR} shows a hidden Markov model (HMM) that, on the
one hand, is patently quite structured, and, on the other, generates a process
with a flat power spectrum \cite{Riec15a}. Indeed, most stochastic processes
seem to hide information about their temporal dependencies in higher-order
statistics \cite{John10a, Jame10a}. Second, as suggested in Ref.
\cite{Samo2007a}, our determination of whether or not a process has long memory
ideally should be invariant under invertible transformations of one's
measurement values. The challenge is not only to find a new statistic that
addresses these two concerns, but to find a statistic that is also easy to
operationalize.

References \cite{Crut92c,Bial00a,Crut01a} suggested a process might be said to
have long memory when the mutual information between its past and future
(excess entropy) diverges, and Ref. \cite{Crut92c} suggested that long memory
is associated with divergent statistical complexity with the effective memory
architecture given by a process' \eM.  By construction, these statistics are
invariant under invertible transformations of the data; and with sufficiently
clever entropy estimation techniques, these statistics are also calculable
directly from time series data.

Unfortunately, there is a paucity of concrete examples upon which to build
intuition as to how these higher-order statistics and the more commonly used
second-order statistics relate.  In part, this lack of concrete examples might
owe somewhat to the fact that it is nontrivial to construct ergodic stationary
processes with divergent excess entropy, though see Refs. \cite{Trav11b,
Debo12a}. (Note that the processes considered in Ref. \cite{Bial00a} were
nonergodic \cite{Crut15a}.)

To that end, we study a tractable class of processes that can have both
divergent excess entropy and Hurst exponent greater than $1/2$: the
\emph{fractal renewal processes} \cite{Smit58a,Gerstner,Beichelt,Barbu} in
which interevent intervals are drawn independently and identically (IID) from a
probability distribution with tails $\propto t^{-\alpha}$. These processes are
very widely used in the physical, biological, and social sciences to model
diverse long-memory phenomena, ranging from current fluctuations in electronic
devices and neuronal spike trains to earthquakes and astrophysical time series
\cite{Lowe93b,Thurner1997,Caki06a,Bian07a,Li08a,Akim10a,Kell12a,Mont13a,Bolo13a,Onag14a}.


Previous studies analyzed the second-order statistics of such processes in some
detail \cite{Lowe93a, daley1999hurst}.  Here, we use techniques inspired by
those in Refs. \cite{Trav11a, Debo12a} to calculate the excess entropy and
statistical complexity of fractal renewal processes for the first time.  We
find that fractal renewal processes have divergent excess entropy only and
exactly when $\alpha=1$ and divergent statistical complexity as $\alpha \to 1$
from above and for all $0 < \alpha < 1$. However, fractal renewal processes
have power-law power spectra for all $0 < \alpha < 2$ \cite{Lowe93a} and Hurst
exponents greater than $1/2$ \cite{daley1999hurst}---the latter being two of
the conventional second-order statistical signatures of ``long memory''. Thus,
even for these relatively straightforward processes, the excess entropy and
statistical complexity encapsulate a different notion of long memory than one
gleans using only second-order statistics. These results also add fractal
renewal processes to a very short list of known stationary ergodic processes
with divergent excess entropy \cite{Trav11a, Debo12a} and so, we hope, they
pave the way for more general comparisons between different definitions of long
memory.

Section \ref{sec:Background} briefly reviews definitions of memory in
stochastic processes. Section \ref{sec:FRPs} calculates informational measures
of memory for fractal renewal processes. Section \ref{sec:Conclusion} then
compares our findings to the second-order statistics calculated by Refs.
\cite{Lowe93a,daley1999hurst} and draws out the lessons for the above
application examples. We close by reflecting on structural organization
associated with long memory.

\vspace{-0.4in}
\section{Background}
\label{sec:Background}
\vspace{-0.1in}

There are many definitions for a stochastic process to have long memory; Ref.
\cite{Samo2007a} provides a particularly helpful survey. Consider a sequence of
$\ell$ observations $\ms_0, \ms_1, \ldots, \ms_{\ell-1}$, realizations of
discrete-valued random variables $\MS_0, \MS_1, \ldots, \MS_{\ell-1}$. For
instance, if the \emph{autocorrelation function} $\mathcal{C}(\tau)$ is
asymptotically a power law multiplied by a slowly varying function $g(\tau)$,
then a process can be said to have ``long memory'':
\begin{align*}
  \mathcal{C}(\tau) & = \sigma^{-2} \sum_{j=0}^\ell (\ms_j - \mu) (\ms_{j+\tau} - \mu) \\
                    & \propto g(\tau) \tau^{-\gamma}
  ~,
\end{align*}
with $0<\gamma <1$, mean $\mu$, and variance $\sigma^2$.
Yet other definitions are based on the decay of the \emph{spectral density}:
\begin{align*}
  \mathcal{P}(f) & = \ell^{-1}
    \left| \sum_{j=0}^\ell \ms_j e^{-i j f} \right|^2.
\end{align*}
The process has long memory when $\mathcal{P}(f)\propto f^{-\beta}L_1(f)$ as $f$ approaches $0$ (where $L_1(f)$ is a slowly varying function near $f=0$) with $0<\beta<1$.
Other definitions still are based on how variances deviate from time-local linear extrapolation.
Starting with the variance of partial sums $S_j = \MS_1 + \cdots + \MS_j$,
one uses the \emph{rescaled-range} statistics:
\begin{align*}
  RS(\ell) & = \frac{\max_{0 \leq j \leq \ell} (S_j - \tfrac{j}{\ell} S_\ell)
  - \min_{0 \leq j \leq \ell} (S_j - \tfrac{j}{\ell} S_\ell)}{\sigma} \\
  & \propto \ell^{-H}
  ~,
\end{align*}
where $H \in (0,1)$ is the \emph{Hurst index}. Processes with $H > 1/2$ are
interpreted as having long memory. Unfortunately, even these second-order
statistics are not always equivalent signatures of long memory. Section 5 of
Ref. \cite{Samo2007a} provides examples of inconsistencies.

In a search for general principles from ergodic theory, Sec. 4 of Ref.
\cite{Samo2007a} proposed that we require a definition of long memory
independent of invertible transformations of the data. That is, if an
invertible transformation is applied pointwise to each observation $\MS_i$, we
would hope that the resulting process has long memory if and only if the
original process had long memory. This desideratum is not satisfied by
definitions based on the above second-order statistics. 

Since strongly mixing processes have short memory and nonergodic processes
could be said to have infinite memory \cite{Crut15a}, Ref. \cite{Samo2007a}
proposed that one or another type of nonmixing property is a good candidate for
long memory in ergodic stationary processes. This criterion satisfies the
invariance desideratum above but can be rather difficult to evaluate.

Fortunately, the information-theoretic notions of memory we consider also
satisfy the transformation-invariant desideratum and have been
successfully deployed
as quantifications for the ``complexity'' of
stochastic processes \cite{Crut88a, Bial00a}. We study
two: the \emph{excess entropy} $\EE = I[\Past;\Future]$, or the mutual
information between a process' past $\Past = \ldots \MS_{-3} \MS_{-2} \MS_{-1}$
and future $\Future = \MS_0 \MS_1 \MS_2 \ldots$
\cite{Crut01a}; and the \emph{statistical complexity} $\Cmu$, or the
amount of information from the past $\Past$ required to exactly predict the future $\Future$
\cite{Crut88a}. When the excess entropy diverges, we are interested in the
asymptotic rate of divergence of finite-length excess entropy estimates
$\EE(\ell)
= I[\Past;\FutureL]$ \cite{Bial00a, Crut01a}. This asymptotic rate of
divergence is also invariant to temporally local convolutions and invertible
transformations of the data \cite{Bial00a}.

To more precisely define and calculate the statistical complexity and the
excess entropy, we need to recall the causal states of computational mechanics.
Consider clustering pasts according to an equivalence relation $\sim$ in which
two pasts are equivalent when they have the same conditional probability
distribution over futures: $\past \sim \past'$ if and
only if $\Prob(\Future|\Past=\past) = \Prob(\Future|\Past=\past')$.  The
resulting clusters are \emph{forward-time causal states} $\St^+$,
which inherit a probability distribution from the probability
distribution over pasts. The \emph{forward-time statistical
complexity} is the entropy of these causal states: $\Cmu^+ =
H[\St^+]$. For more detail, see Refs. \cite{Shal98a,Lohr09a}.
 
We can similarly define the \emph{reverse-time causal states} $\St^-$ by
clustering futures with equivalent conditional probability distributions over
pasts: $\future \sim \future'$ if and only if $\Prob(\Past|\Future=\future) =
\Prob(\Past|\Future=\future')$. The \emph{reverse-time statistical complexity} is the entropy of those reverse-time causal states: $\Cmu^- = H[\St^-]$.
Renewal processes are time-reversal invariant \cite{Marz14b}, or \emph{causally
reversible}, so throughout the following we denote the statistical complexity as $\Cmu = \Cmu^+ = \Cmu^-$ without loss of precision.

Reverse-time causal states and forward-time causal states can be used to calculate the excess entropy \cite{Crut08a, Crut08b}:
\begin{align*}
\EE = I[\St^+;\St^-]
  ~.
\end{align*}
For discrete-time processes, $\EE$ is a lower bound on $\Cmu$:
\begin{align}
\EE \leq \Cmu
  ~.
\label{eq:DTEEvCmu}
\end{align}
In other words, for discrete-time processes, if statistical complexity is
finite, then so is excess entropy. Conversely, if excess entropy is infinite,
then statistical complexity is infinite.

Often continuous-time processes have an uncountable set of causal states.
For them, the statistical complexity is taken to be the differential entropy:
\begin{align*}
\CCT & = H[\St^+] \\
     & = - \int_{\Delta} d \mu (\st^+) \log \mu (\st^+)
  ~,
\end{align*}
where we have the simplex $\Delta$ of causal states and $\mu(\st^+)$ is their
measure in $\Delta$. In the continuous-time setting, the inequality analogous
to Eq. (\ref{eq:DTEEvCmu}) no longer necessarily holds \cite{Marz14e}. We call
the differential entropy $\CCT$ the \emph{continuous-time statistical
complexity} to distinguish it from the discrete-time statistical complexity
$\Cmu$, but simply refer to it as the statistical complexity when context is
clear.
 
One can also define finite-time reverse-time causal states, denoted
$\St^-_\ell$, by clustering futures of finite-length $\ell$ with the same
equivalence relation as above.  From these, we obtain finite-length
reverse-time statistical complexity $\Cmu^{-_\ell}=H[\St^-_\ell]$,
respectively.  These can be used to calculate finite-future excess entropy
estimates: $\EE(\ell) = I[\St^+;\St^-_\ell]$ \cite{Crut08a,Crut08b}.

For discrete-alphabet, discrete-time processes, the statistical complexity is
invariant to relabelings of the measurement alphabet. However, as just noted,
when the causal states are uncountable, the statistical complexity involves a
differential entropy, and differential entropies are not invariant to
invertible transformations of the coordinate system of the distribution's
support. A prosaic example of this is given in Ref. \cite{Marz14a}. Modulo such
factors, whether or not statistical complexity diverges, the rate of divergence
of its finite-length estimates $\Cmu^{\ell}$ is invariant to temporally local
convolutions of the data.

Realizations from a renewal process consist of sequences of events separated by
epochs of quiescence, the lengths of which are drawn independently from the
same \emph{interevent} distribution. Throughout, when discussing a
discrete-time renewal process, we use the following notation \cite{Marz14b}:
$F(n)$ is the \emph{interevent count} probability distribution function; $w(n)
= \sum_{n'}^\infty F(n')$ is the \emph{survival function}; and $\mu$ is its
\emph{mean interevent count}. We use the following notation for continuous-time
renewal processes: $\phi(t)$ is the \emph{waiting time} distribution; $\Phi(t)$
is its \emph{survival function}; and $\MISI$ is the \emph{mean interevent
interval}. \emph{Fractal renewal processes} have survival functions that have
power-law tails, as introduced shortly.

\section{Intrinsic Memory in Fractal Renewal Processes}
\label{sec:FRPs}

Fractal renewal processes---those with power-law interevent interval
probability density functions---can have long memory in the sense of Ref.
\cite{graves2014brief}. For instance, they can have Hurst index $H > 1/2$
\cite{daley1999hurst} and their autocorrelation function can be
(asymptotically) a power law \cite{Lowe93a}.  Fractal renewal processes have
been implicated in a variety of complex natural processes, to which the
introduction alluded. Might these processes also have infinite statistical
complexity or infinite excess entropy? To the best of our knowledge, the excess
entropy and statistical complexity of fractal renewal processes have yet to be calculated.

Calculating statistical complexity and excess entropy can be challenging when
going beyond finite causal-state processes \cite{Crut13a}. To make progress
with bounding the excess entropy of fractal renewal processes, we use two
tools. The first tool is to coarse grain by time-binning. The Data Processing
Inequality \cite{Cove06a} then implies that the excess entropy of a
discrete-time renewal process is always upper-bounded by the excess entropy of
the corresponding continuous-time renewal process.  See App.~\ref{app:time_bin}. The second tool allows us to
calculate excess entropy and statistical complexity even when the mean rate of
events vanishes by conditioning on the presence of a proxy event.  This tool
was inspired by previous work \cite{Trav11b} and is summarized in App.~\ref{app:condition}.

Fractal renewal processes are typically considered in continuous-time, with
interevent intervals generated independently and identically distributed (IID)
from the probability density function:
\begin{align}
\phi(t) = \begin{cases} 0 & t<1 \\  \alpha t^{-(\alpha+1)} & t\geq 1 \end{cases}
  ~.
\label{eq:phi}
\end{align}
The probability of seeing an interevent interval of length $t$ or larger is
the survival function:
\begin{align}
\Phi(t) & = \int_t^{\infty} \phi(t') dt' \nonumber \\
  & = \begin{cases} 1 & t<1 \\ t^{-\alpha} & t\geq 1
  \end{cases}
  ~.
\label{eq:Phi}
\end{align}
Time intervals are given in terms of the shortest possible interevent
interval. When $\alpha>1$, the mean interevent interval $\MISI =
\frac{\alpha}{\alpha-1}$ is finite; when $0 < \alpha \leq 1$, the mean
interevent interval is infinite, but one will always eventually see an event.

Appendix \ref{sec:FRP} describes how to manipulate the continuous-time analog
of Eq.~(\ref{eq:EEL2}) to obtain:
\begin{align}
\ECT & = \begin{cases}
  \log \frac{\alpha^2}{\alpha-1}-1 & \alpha>1 \\
  \infty & \alpha=1 \\
  \frac{\alpha^2+\alpha-1}{\alpha(1-\alpha)}
  + \log\frac{\alpha}{1-\alpha}-(1-\alpha) K_{\alpha} & \alpha<1
  \end{cases}
  ~,
\label{eq:EE_FRP}
\end{align}
where $K_{\alpha} = \int_0^{\infty} (u^{-\alpha}-(1+u)^{-\alpha})\log
(u^{-\alpha}-(1+u)^{-\alpha}) du$. Note that at small values of $\alpha$,
$K_{\alpha}$ is difficult to evaluate numerically due to the integrand's long
tails, even when $\ECT$ is quite small. For instance, when $\alpha=1/4$, $\ECT
\approx 0.089$ nats, but $\int_0^{N} (u^{-\alpha}-(1+u)^{-\alpha})\log
(u^{-\alpha}-(1+u)^{-\alpha}) du$ does not return positive estimates for the
excess entropy until $N\geq 10^{11}$. A more obvious benefit of
Eq.~(\ref{eq:EE_FRP}), then, is that we can study the excess entropy's
asymptotic behavior near $\alpha=1$, where $\ECT(\ell) \sim \log\log \ell$. This
divergence is slower than any previously reported divergence \cite{Bial00a,
Trav11b, Debo12a}, but is a divergence nonetheless.

When $\alpha>1$ but close to its critical value, the excess entropy diverges as $\sim \log\frac{1}{\alpha-1}$. As $\alpha\rightarrow\infty$, $\ECT$ diverges as
$\log\alpha$.  This point is discussed more fully later on.

The discrete-time analog of fractal renewal processes has a survival function:
\begin{align}
w(n) = \begin{cases} 1 & n=0 \\ n^{-\alpha} & n \geq 1 \end{cases}
  ~.
\label{eq:DTFRP_wofN}
\end{align}
The transient (small $n$) behavior of $w(n)$ may not match that in some
applications, but only $w(n)$'s asymptotic behavior is relevant to $\EE$'s
divergence. Moreover, App.~\ref{app:time_bin} guarantees that $\EE$ is finite when
$\alpha\neq 1$ and that at $\alpha=1$ its divergence is at most $\log\log
\ell$. Additional arguments in App.~\ref{sec:FRP}, in turn, show that $\EE(\ell)$ diverges at $\alpha=1$ as $\log\log \ell$.


The excess entropy $\EE$ captures the amount of predictable randomness of a
stochastic process. As a comparison, we are also interested in the statistical
complexity $\Cmu$ of discrete-time and continuous-time fractal renewal
processes. The statistical complexity is the number of bits required to
losslessly predict ($\EE$ nats of) the process' future. Sometimes, $\Cmu$ is
not much larger than $\EE$; for discrete-time periodic processes, the two are
equivalent and equal to the logarithm of the period. More often than not,
$\Cmu$ is infinite while $\EE$ is finite; e.g., for processes generated by most
(nonunifilar) Hidden Markov Models.

\emph{Cryptic processes} have large statistical complexity and small excess
entropy \cite{Crut08a}; the larger the crypticity, the more that a process' true structure is ``hidden'' from the observer. An open question is whether or not fractal renewal processes, with their statistical signatures of complexity, are highly cryptic. So, we focus some attention now on evaluating $\Cmu$ for fractal renewal processes.

We can calculate $\Cmu$ of time-binned continuous-time renewal
processes in the infinitesimal-$\tau$ limit \cite{Marz14e}:
\begin{align*}
{\Cmu}_{\tau} & \sim \log \frac{1}{\tau}
  - \int_0^{\infty} \frac{\Phi(t)}{\MISI} \log \frac{\Phi(t)}{\MISI} dt
  ~.
\end{align*}
As we will discuss elsewhere, the above expression is the differential entropy
over continuous-time causal states---the expression given in
Sec.~\ref{sec:Background} as the ``continuous-time statistical complexity''
$\CCT$---plus the logarithm of our time-bin resolution. Thus, ${\Cmu}_{\tau}$'s
$\log \tfrac{1}{\tau}$ divergence is an artifact of our failure to use the
differential entropy when calculating memory storage requirements of continuous
random variables \cite{Cove06a}. As a result, we focus on ${\Cmu}_{\tau}$'s
nondivergent component, $\CCT = \lim_{\tau\rightarrow 0} \left( {\Cmu}_{\tau} +
\log \tau \right)$, or what was earlier called the continuous-time statistical
complexity. Straightforward algebra shows that:
\begin{align}
\CCT &= \begin{cases}
  \frac{1}{\alpha-1}+ \log\frac{\alpha}{\alpha-1} & \alpha>1 \\
  \infty & \alpha\leq 1
  \end{cases}
  ~.
\label{eq:Cmu_FRP}
\end{align}
Again, we can say that the (continuous-time) $\Cmu$ diverges whenever the mean
interevent interval $\MISI$ diverges. When $\alpha\leq 1$, finite-length
statistical complexity estimates adapted to the continuous-time case from
Eq.~(\ref{eq:CmuL}) diverge as:
\begin{align*}
\Cmu^{+_\ell}\sim \begin{cases}
   \log \ell & \alpha<1 \\ \frac{1}{2}\log \ell & \alpha=1
   \end{cases}
  ~.
\end{align*}
So, the special nature of $\alpha=1$ is also revealed as a discontinuity in
rates of divergence of the finite-length statistical complexity. In particular,
the least cryptic fractal renewal process, among fractal renewal processes with
divergent statistical complexity, is the process generated when $\alpha=1$.

\begin{figure}[h!]
\includegraphics[width=0.5\textwidth]{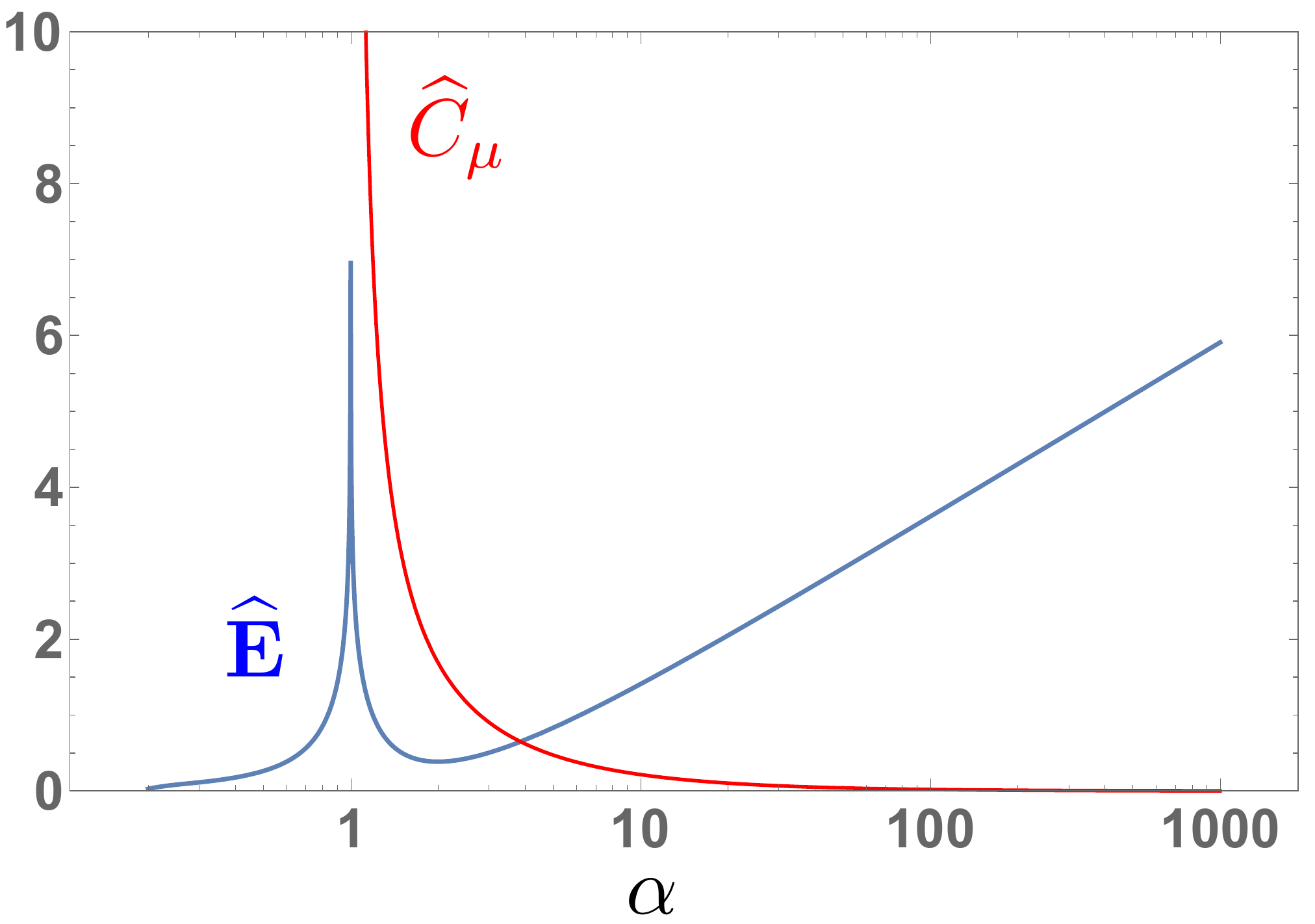}
\caption{\textbf{Excess entropy $\widehat{\EE}$ and statistical complexity
  $\widehat{\Cmu}$ of continuous-time fractal renewal processes:} Process
  realizations are generated by drawing interevent intervals IID from the
  probability density function $\phi(t) = \alpha t^{-(\alpha+1)}$ for $t\geq 1$
  and $0$ otherwise. $\ECT$ in nats as a function of $\alpha$, evaluated using
  Eq.~(\ref{eq:EE_FRP}). The nondivergent component of statistical complexity
  $\CCT$ in nats as a function of $\alpha$, evaluated using
  Eq.~(\ref{eq:Cmu_FRP}). Note that $\CCT$ is a differential entropy and so not
  necessarily larger than the excess entropy $\ECT$; a subtlety when working
  with continuous-time processes.
  }
\label{fig:FRP}
\end{figure}

Equations~(\ref{eq:EE_FRP}) and (\ref{eq:Cmu_FRP}) are plotted in
Fig.~\ref{fig:FRP}. The divergences in $\ECT$ and $\CCT$ at $\alpha=1$ are
apparent in the plot. If $\ECT$ and $\CCT$ are taken to be systems-agnostic
order parameters, then a fractal renewal process exhibits a nonequilibrium
phase transition exactly when its mean interevent interval diverges.

The behavior of $\ECT$ and $\CCT$ as $\alpha$ tends to infinity also deserves
special mention, as the process appears to become infinitely predictable
($\ECT\rightarrow\infty$) while requiring less memory for prediction
($\CCT\rightarrow 0$). As $\alpha$ tends to $\infty$, $\phi(t)$ becomes more
and more sharply peaked at $t=1$. In other words, the process moves closer and
closer to that of a periodic process with period $1$. Periodic processes are
random enough, in that the phase of the process could be any real number
between $0$ and the period. In the language of computational mechanics, the
causal state \textit{is} the phase, and its differential entropy---the
continuous-time statistical complexity $\CCT$---is the logarithm of the
process' period. As $\alpha\rightarrow\infty$, the mean interevent interval
$\MISI = \frac{\alpha}{\alpha-1}$ tends to $1$, and the continuous-time
statistical complexity correspondingly tends to $\log 1 = 0$. However, periodic
processes are also highly predictable, in that the time to next event is
determined by the time since last event; hence, the differential entropy of the
time to next event \textit{conditioned on} the time since last event tends
towards negative infinity, resulting in an infinite $\ECT = \CCT -
H[\St^-|\St^+] \rightarrow \infty$. Similar behavior was seen in
Ref.~\cite{Marz14e} as the noisiness of spike trains tended to zero, though.
The least cryptic fractal renewal process, then, occurs in the limit that
$\alpha$ tends to infinity.

\section{Conclusion}
\label{sec:Conclusion}

We showed that a fractal renewal process's excess entropy diverges precisely
when its mean interevent interval diverges. This adds a relatively easily
understood process and one of much broader applicability to the existing list of ergodic stationary processes with divergent excess entropy \cite{Trav11b, Debo12a}.


Notably, the expected number of events observed in a finite time interval for a
fractal renewal process with divergent excess entropy is zero. This brings in
an interpretational challenge. A process that, on average, produces arbitrarily long silence is
not often described as random. So, should not the excess entropy of a point process with infinite
mean interevent interval be zero? However, the mutual information between
finite-length pasts and futures, assuming that we do see an event, can diverge.
And, we will almost surely see an event when we view a semi-infinite past.

Our calculations revealed that fractal renewal processes flip from finite to
divergent statistical complexity and exhibit divergent excess entropy exactly
when the mean interevent interval diverges. These information-theoretic
measures of memory point to the power-law coefficient $\alpha=1$ as being a
``critical'' parameter in this process family. When the mean interevent
interval is finite, both excess entropy and continuous-time statistical
complexity are finite, though excess entropy grows unbounded as $\alpha$ tends
to infinity. When the mean interevent interval is infinite and the power-law
coefficient is not $\alpha=1$, excess entropy is finite, but continuous-time
statistical complexity is infinite.

Employing signatures of long memory based on second-order statistics suggests,
instead, that $\alpha=2$ was a ``critical point''. Specifically, the power
spectrum of a fractal renewal process exhibits power-law scaling when $\alpha<
2$ \cite{Lowe93a}, and the Hurst index of the processes with $\alpha<2$ is
greater than $1/2$ and increases with decreasing $\alpha$
\cite{daley1999hurst}. Therefore, at a minimum, drawing conclusions about a
process' complex organization via such low-order statistics can be ambiguous.

Finally, our results suggest that certain previously studied experimental
phenomenon are poised at a critical point between finite and infinite
``memory'', as suggested by many others using other definitions of criticality
\cite{mora2011biological}. The stochastic process of neuron membrane ion
channels opening and closing has divergent excess entropy when the kinetic rate
adopts the form $k_\text{eff}(t)\approx t^{-1}$. This may be the case for some
potassium-selective channels in cultured mouse hippocampal pyramidal cells near
resting membrane voltage, $V=-60~\text{V}$ \cite[Fig. $10$, bottom
right]{liebovitch1987fractal}. Similarly, the phenomenological fit of the
stopping probabilities used for Wikipedia edit-revert time series has divergent
statistical complexity when $\alpha=1$ and divergent excess entropy when $p=1$
as well \cite{Dedeo13,Dedeo14}. This seems to suggest that increased
cooperativity between editors drives Wikipedia towards increasing its social
memory.

However, one lesson from our results is tantamount to a cautionary note on
interpreting the implicated memory organization. To the extent that the
estimated fractal renewal processes with divergent memory are good models, one
cannot conclude that the content of that memory reflects sophisticated
computational processing or highly organized storage of detailed information.
Indeed, like all renewal processes, fractal renewal processes are simple: they
count up to some threshold and reset. Surely these coarse statistics, while
useful and even necessary as tools for a first-cut analysis, fall far short of
fully describing the hierarchies of information processing in neurons and the
rich social dynamics driving Wikipedia's accumulating human knowledge.

To close, let's return to our initial discussion of statistical signatures of
structural organization. We drew a comparison of divergent memory in ergodic
processes to that we previously identified in the so-called Bandit nonergodic
processes \cite{Crut15a}. The mechanism underlying the latter was rather
straightforward: from trial to trial the process remembers the operant ergodic
component subprocess and so uses an infinite memory and exhibits an excess
entropy that diverges as $\log \ell$. The case for ergodic process is more
subtle. For renewal processes we showed that the divergence is $\log \log
\ell$. What's the associated mechanism? Renewal processes track time between
events and so, in computational model terms, it appears that the process
somehow embeds a counter \cite[Sec. 4.5.2]{Crut92c}. An interesting contrast is
the $\log \ell$ excess entropy divergence seen at the onset of chaos through
period-doubling, associated with pushdown stack mechanism \cite[Sec.
4.5.1]{Crut92c}, and seen in the branching copy process \cite{Trav11b}. At this
stage, though, the possibility of unique associations between the form of
information measure divergence and mechanism is not sufficiently well explored.
Nonetheless, with further extension and refinement information measures and
their divergences will become increasingly more insightful diagnostics of
nature's diverse forms of intrinsic computation.

\acknowledgments

The authors thank W. Bialek, S. Dedeo, and P. Riechers for helpful discussions and the Santa Fe Institute and the City University of New York for hospitality during visits. This material is based upon work supported by, or in part by, the U.S. Army Research Laboratory and the U. S. Army Research Office under contracts W911NF-13-1-0390, W911NF-13-1-0340, and W911NF-12-1-0288. S.E.M. was funded by a National Science Foundation Graduate Student Research Fellowship and the U.C. Berkeley Chancellor's Fellowship.

\appendix

\section{Continuous- versus Discrete-time Excess Entropies}
\label{app:time_bin}

%
%

Often, integrals are easier to evaluate than the corresponding sums.  One
practical goal, leveraging this below, is to relate the excess entropy of
time-binned continuous-time processes to that of corresponding discrete-time renewal processes.

Reference \cite{Marz14b} found that the excess entropy of a discrete-time
renewal process is:
\begin{align}
\EE = \log (\mu+1) & - \frac{2}{\mu+1} \sum_{n=0}^{\infty} w(n)\log w(n) 
  \nonumber \\
  & + \frac{1}{\mu+1} \sum_{n=0}^{\infty} (n+1)F(n)\log F(n)
  ~.
\label{eq:EEDT}
\end{align}
While Ref. \cite{Marz14e} showed that the excess entropy of a
continuous-time renewal process $\MS(t)$ is:
\begin{align}
\ECT & = \I[\MS(t)_{t<0} ; \MS(t)_{t\geq0}] \nonumber \\
  & = \log \MISI - \frac{2}{\MISI} \int_0^{\infty} \Phi(t)\log \Phi(t) dt 
  \nonumber \\
  & \quad\quad\quad\quad + \frac{1}{\MISI} \int_0^{\infty} t\phi(t)\log \phi(t) dt
  ~,
\label{eq:EE}
\end{align}
which is in units of \emph{nats} when the mean interevent interval $\MISI$ is
finite.

Consider time-binning the continuous-time point process $\MS(t)$ by asking how
many events are observed in an interval $[t,t+\tau)$. If at least one event is
observed, then we record a $1$; if no events are observed, then we record a
$0$. This data labeling technique is common; e.g., when studying neural spike
trains. The probability of observing at least $n$ counts between successive
$1$s is given by:
\begin{align*}
w_{\tau}(n) = \Phi(n\tau)
  ~.
\end{align*}
When $\tau=1$, then the survival function of the time-binned process is exactly
that of the discrete-time renewal process with excess entropy given in
Eq.~(\ref{eq:EEDT}).

The excess entropy or estimates thereof for a discrete-time renewal process are
upper bounded by the excess entropy of a corresponding continuous-time renewal
process, as shown shortly. This is a special case
of a more general statement: coarse-graining a time series always reduces its
excess entropy, due to the Data Processing Inequality. This statement can be easily generalized to other discrete-alphabet, continuous-time processes.
Despite its simplicity, it proves useful for the calculations to come in Sec.~\ref{sec:FRPs}.

In particular, let $\ECT$ denote the excess entropy of a continuous-time
renewal process $\MS(t)$ with survival function $\Phi(t)$ and $\EE$ the excess
entropy of the discrete-time renewal process $\MS_t$ with survival function
$w(n) = \Phi(n)$ for all nonnegative integers $n$. Then, when $\ECT < \infty$:
\begin{align*}
\EE & \leq \ECT
  ~.
\end{align*}
To see this, let $\EE_{\tau}$ denote the excess entropy of the discrete-time
process that comes from time-binning the continuous-time renewal process with
discretization bin size $\tau$. To obtain the above inequality, we apply the
Data Processing Inequality:
\begin{align*}
\EE_{1/n} & = I[\ldots,\MS(-2/n),\MS(-1/n);\MS(0),\MS(1/n),\ldots] \\
&\geq I[\ldots,\MS_{-2},\MS_{-1};\MS_0,\MS_1,\ldots] \\
&= \EE_{1}
  ~.
\end{align*}
If we take the limit of the left-hand side as $n\rightarrow\infty$, we obtain:
\begin{align*}
\EE_{\tau=1} & \leq \lim_{n\rightarrow\infty} \EE_{1/n} \\
             & = \lim_{\tau\rightarrow 0} \EE_{\tau}
  ~.
\end{align*}
Again by the Data Processing Inequality, $\EE_{\tau=1}$ is lower-bounded by the
mutual information between the counts since last event and counts to next
event, as the former is a function of the past and the latter is a function of
the future: $\EE \leq \EE_{\tau=1}$. By definition \cite{pinsker1960information},
$\lim_{\tau\rightarrow 0} \EE_{\tau}=\ECT$.


\section{Renewal Processes with Infinite Mean Intervent Intervals}
\label{app:condition}

When the mean interevent interval $\MISI$ (or $\mu$) is infinite, the formulae
for excess entropy in Eqs.~(\ref{eq:EEDT}) and (\ref{eq:EE}) no longer apply.
Causal states, however, still provide a useful framework for calculating it.
Using them we introduce an analysis method for discrete-time renewal processes
in this case. The obvious extensions to continuous-time renewal processes
follow when we replace $F(n)$ with $\phi(t)$, $w(n)$ with $\Phi(t)$, and
summations with integrals.

We calculate $\EE(\ell)$ for renewal processes with infinite $\mu$ via an
analysis technique inspired by Ref. \cite{Trav11b} and then calculate $\EE$ as
a limit of $\EE(\ell)$ as $\ell$ tends to infinity, seemingly valid for ergodic processes. First, we would like to
directly calculate $\EE(\ell)$ in terms of forward and reverse-time causal
states \cite{Crut08a}: $\EE(\ell) = I[\Past;\FutureL] =  I[\St^+;\St^-_\ell]$,
where $\St^-_\ell$ are finite-time reverse-time causal states. Unfortunately,
inspecting the corresponding joint probability distribution in App. II of Ref.
\cite{Marz14b} shows that while we can identify the joint probability
distribution up to a normalization constant, this normalization constant is
infinite when $\mu$ is infinite.

So, we define a ``proxy'' binary random variable $U_\ell$ which is $1$ if there
has been an event sometime in $\FutureL$ and past $\Past$, and $0$ otherwise.
A little reflection shows that $\Prob(U_\ell=0) = \lim_{N\rightarrow\infty}
w(N+\ell)
= 0$.  Even so, this auxiliary random variable is a surprisingly useful
construct.  A standard information-theoretic decomposition gives $\EE(\ell) =
I[\St^+;\St^{-_\ell}|U_\ell] + I[\St^+;\St^-_\ell;U_\ell]$, but since
$\Prob(U_\ell=0) = 0$,
we have that $I[\St^+;\St^-_\ell|U_\ell] = I[\St^+;\St^-_\ell|U_\ell=1]$ and
$I[\St^+;\St^-_\ell;U_\ell]=0$. Altogether this yields:
\begin{align*}
\EE(\ell) & = I[\St^+;\St^-_\ell|U_\ell=1]
  ~.
\end{align*}
The conditional probability distribution $\Prob(\St^+,\St^-_\ell|U_\ell=1)$ is
normalizable and, as shown in App.~\ref{sec:Calculation1}, leads to:
\begin{align}
\EE(\ell) & = \log Z(\ell) - \frac{1}{Z(\ell)} \sum_{n=0}^\ell w(n)\log w(n) \nonumber \\
& \quad\quad - \frac{1}{Z(\ell)}\Big( \sum_{n=0}^{\infty} (w(n)-w(n+\ell+1))\nonumber \\
& \quad\quad\quad \times \log (w(n)-w(n+\ell+1))\Big) \nonumber \\
& \quad\quad + \frac{1}{Z(\ell)} \sum_{n=0}^\ell (n+1) F(n)\log F(n) \nonumber \\
& \quad\quad + \frac{\ell+1}{Z(\ell)} \sum_{n=\ell+1}^{\infty} F(n)\log F(n)
  ~,
\label{eq:EEL2}
\end{align}
where $Z(\ell) = \sum_{n=0}^\ell w(n)$.
If $\lim_{\ell \rightarrow\infty} \EE(\ell)$ diverges, then we look for the
asymptotic rate of divergence of $\EE(\ell)$. Otherwise, the process' excess
entropy can be defined as $\EE = \lim_{\ell \rightarrow\infty} \EE(\ell)$. We expect $\EE$ will often be finite even when $\mu$ diverges.

A similar method allows us to calculate $\Cmu$ when mean interevent count is
infinite. This time, we define $U_\ell$ as a proxy random variable that is $1$ if
there has been an event in $\PastL$ and $0$ otherwise. Since $U_\ell$ is a
function of $\St^+$, a standard information-theoretic identity implies that:
\begin{align*}
\Cmu = \H[\St^+|U_\ell] + \H[U_\ell]
\end{align*}
and, in particular:
\begin{align*}
\Cmu = \lim_{\ell \rightarrow\infty} \left( \H[\St^+|U_\ell] + \H[U_\ell] \right)
  ~.
\end{align*}
As before, $\lim_{\ell \rightarrow\infty} \Prob(U_\ell =0) =
\lim_{\ell \rightarrow\infty}
w(\ell) = 0$, so $\lim_{\ell \rightarrow\infty} \H[U_\ell]=0$.  Also,
$\H[\St^+|U_\ell] =
\Prob(U_\ell=0) \H[\St^+|U_\ell=0]+\Prob(U_\ell=1) \H[\St^+|U_\ell=1]$ by definition. Since
there is only one semi-infinite past without an event,
$\lim_{\ell \rightarrow\infty} \H[\St^+|U_\ell =0] = 0$. And,
$\H[\St^+|U_\ell=1] =
-\sum_{n=0}^\ell \frac{w(n)}{Z(\ell)} \log \frac{w(n)}{Z(\ell)}$. Altogether, this
implies:
\begin{align}
\Cmu = \lim_{\ell \to \infty} \sum_{n=0}^\ell \frac{w(n)}{Z(\ell)}
   \log \left(1/\frac{w(n)}{Z(\ell)}\right)
  ~.
\label{eq:CmuL}
\end{align}
One can also study the growth rate of finite-time statistical complexity
estimates which are, after a moment's reflection, the $\Cmu^{\ell} = -
\sum_{n=0}^\ell \frac{w(n)}{Z(\ell)} \log \frac{w(n)}{Z(\ell)}$ estimates
above in Eq.~(\ref{eq:CmuL}).

One comment, perhaps obvious from Eqs.~(\ref{eq:EEL2}) and (\ref{eq:CmuL}), is
that whether or not $\EE$ and $\Cmu$ diverge depends entirely on the asymptotic
form of $F(n)$. Another is that the sums in Eq.~(\ref{eq:EEL2}) can be quite
difficult to evaluate numerically when the renewal process has long-range
temporal correlations, since then $F(n)$ decays slowly with $n$.

\section{Finite-time Excess Entropy Estimates with Infinite Mean Interevent Interval}
\label{sec:Calculation1}

From App. II of Ref. \cite{Marz14b}:
\begin{align*}
\Prob(\St^+=\st^+, & \St^-_\ell = \st^-|U_\ell=1) \\
   & = \frac{1}{Z}
   \begin{cases} F(\st^+ + \st^-) & \st^-\leq \ell \\ 0 & \st^- = \ell + 1
   \end{cases}
  ~,
\end{align*}
where the normalization constant is:
\begin{align*}
Z & = \sum_{\st^-=0}^{\ell} \sum_{\st^+=0}^{\infty}F(\st^+ + \st^-) \\
  & = \sum_{\st^-=0}^\ell w(\st^-)
  ~.
\end{align*}
The marginals are easily calculated:
\begin{align*}
\Prob(\St^+=\st^+|U_\ell=1) & = \frac{1}{Z} (w(\st^+)-w(\st^+ + \ell +1))
\end{align*}
and:
\begin{align*}
\Prob(\St^-_\ell =\st^-|U_\ell=1) & = \frac{1}{Z}
  \begin{cases} w(\st^-) & \st^-\leq \ell \\ 0 & \st^- = \ell +1
  \end{cases}
  ~.
\end{align*}
From this, we calculate finite-length excess entropy in nats:
\begin{align*}
\EE(\ell) & = H[\St^-_\ell|U_\ell=1] + H[\St^+|U_\ell=1] \nonumber \\
       & \quad\quad\quad\quad\quad - H[\St^+,\St^-_\ell|U_\ell=1] \\
       & = \log Z - \frac{1}{Z} \sum_{n=0}^\ell w(n)\log w(n) \nonumber \\
       & \quad\quad\quad - \frac{1}{Z} \Big(\sum_{n=0}^{\infty} (w(n)-w(n+\ell+1))\nonumber \\
       & \quad\quad\quad\quad\quad \times \log (w(n)-w(n+\ell+1))\Big) \nonumber \\
       & \quad\quad\quad\quad + \frac{1}{Z} \sum_{n=0}^{\infty} \sum_{m=0}^\ell F(n+m)\log F(n+m)
  ~.
\end{align*}
This simplifies to:
\begin{align*}
\EE(\ell) & = \log Z - \frac{1}{Z} \sum_{n=0}^\ell w(n)\log w(n) \nonumber \\
       & \quad\quad\quad- \frac{1}{Z}\Big( \sum_{n=0}^{\infty} (w(n)-w(n+\ell+1))\nonumber \\
       & \quad\quad\quad\quad\quad \times \log (w(n)-w(n+\ell+1))\Big) \nonumber \\
       & \quad\quad\quad + \frac{1}{Z} \sum_{n=0}\ell (n+1) F(n)\log F(n) \nonumber \\
       & \quad\quad\quad + \frac{\ell+1}{Z} \sum_{n=\ell+1}^{\infty} F(n)\log F(n)
	   ~.
\end{align*}

Similar manipulations hold for continuous-time processes. Briefly, the time since last event $t$ and time to next event $t'$ have a joint probability distribution proportional to $\phi(t+t')$, since the time since last event plus the time to next event is an interevent interval.


\section{Fractal Renewal Processes}
\label{sec:FRP}

The $\alpha>1$ case simply requires substituting $\phi(t)$ and $\Phi(t)$ from
Eqs.~(\ref{eq:phi})-(\ref{eq:Phi}) into Eq.~(\ref{eq:EE}) and solving:
\begin{align}
\ECT &= \log \MISI - \frac{2}{\MISI} \int_0^{\infty} \Phi(t)\log \Phi(t) dt 
  \nonumber \\
  & \quad\quad\quad\quad + \frac{1}{\MISI} \int_0^{\infty} t\phi(t)\log \phi(t) dt.
\label{eq:EERenCT}
\end{align}
After straightforward calculations, we find that:
\begin{align*}
\MISI & = \frac{\alpha}{\alpha-1} ~,\\
\frac{1}{\MISI}\int_0^{\infty} \Phi(t)\log \Phi(t) dt
  & = -\frac{1}{\alpha-1} ~,~\text{and}\\
\frac{1}{\MISI}\int_0^{\infty}t \phi(t)\log \phi(t) dt
  & = \log\alpha-\frac{\alpha+1}{\alpha-1}
  ~.
\end{align*}
These together yield:
\begin{align*}
\ECT = \log \frac{\alpha^2}{\alpha - 1} - 1
  ~.
\end{align*}

Now, we turn our attention to the case of $0<\alpha\leq 1$.
There are two possibilities for $\ECT$ when $0<\alpha\leq 1$. One is that
$\ECT$ diverges, in which case, we only care about the asymptotic rate of
divergence of $\ECT(\ell)$. The other possibility is that $\ECT$ does not diverge,
in which case, we only care about contributions $Q$ to $\ECT(\ell)$ that are not
$o(1)$; i.e., that satisfy $\lim_{\ell \rightarrow\infty} Q \neq 0$. Our strategy
in evaluating $\ECT(\ell)$ from Eq.~(\ref{eq:EERenCT}) is to systematically find
closed-form expressions for all components that are not $o(1)$.

Direct solution gives:
\begin{align}
Z & = \begin{cases}
  \frac{\ell^{1-\alpha}}{1-\alpha} & \alpha<1 \\ \log \ell & \alpha=1
  \end{cases}
  ~,
\label{eq:Q0}
\end{align}
plus components of $o(1)$:
\begin{align}
-\frac{1}{Z} \int_0^\ell \Phi(t)\log \Phi(t) dt
  & = \begin{cases} -\frac{\alpha}{1-\alpha} + \alpha \log \ell & \alpha<1
  \\ \frac{1}{2}\log \ell & \alpha=1
  \end{cases}
\label{eq:Q1}
\end{align}
plus components of $o(1)$; and:
\begin{align}
& \frac{1}{Z} \int_1^{\ell}t\phi(t)\log\phi(t) dt
  + \frac{\ell}{Z}\int_\ell^{\infty} \phi(t)\log \phi(t) dt \nonumber \\
  & = \begin{cases}
  -\frac{1-\alpha-2\alpha^2}{\alpha(1-\alpha)}
  + \log \alpha - (1+\alpha)\log \ell & \alpha<1
  \\ -2 - \log \ell & \alpha=1
  \end{cases}
  ~,
\label{eq:Q2}
\end{align}
plus components of $o(1)$.

Finally, we address the only component with no simple closed-form expression:
\begin{align*}
& \frac{1}{Z}\int_0^{\infty} (\Phi(t)-\Phi(t+\ell))\log
(\Phi(t)-\Phi(t+\ell))dt \\
   & \quad = \frac{1}{Z} \int_1^{\infty} (t^{-\alpha}-(t+\ell)^{-\alpha})
   \log (t^{-\alpha}-(t+\ell)^{-\alpha}) dt \\
   & \quad\quad + \frac{1}{Z}\int_0^1 (1-(t+\ell)^{-\alpha})\log
   (1-(t+\ell)^{-\alpha}) dt
  ~.
\end{align*}
Since:
\begin{align*}
\lim_{\ell\rightarrow\infty} \frac{1}{Z}\int_0^1
(1-(t+\ell)^{-\alpha})\log (1-(t+\ell)^{-\alpha}) dt = 0
  ~,
\end{align*}
we ignore that term as a correction of $o(1)$. The case for $\alpha=1$ can
actually be evaluated explicitly since $\frac{1}{t}-\frac{1}{t+\ell} =
\frac{\ell}{t(t+\ell)}$:
\begin{align*}
\lim_{\ell \rightarrow\infty} \frac{1}{Z}
  \int_1^{\infty} \frac{\ell}{t(t+\ell)} \log (\frac{\ell}{t(t+\ell)}) dt
  = -\frac{1}{2}\log \ell
  ~.
\end{align*}
Now, consider the case of $\alpha<1$. We extract the asymptotic scaling in
$\ell$ of the first term by the change of variables $u=\ell t$, giving:
\begin{align*}
& \frac{1}{Z} \int_1^{\infty}
  (t^{-\alpha}-(t+\ell)^{-\alpha})\log (t^{-\alpha}-(t+\ell)^{-\alpha}) dt \\
  & = \frac{\ell^{1-\alpha}}{Z} \int_{1/\ell}^{\infty} (u^{-\alpha}-(1+u)^{-\alpha}) 
  \log (\ell^{-\alpha}(u^{-\alpha}-(1+u)^{-\alpha})) du \\
  & =  -\alpha \frac{\ell^{1-\alpha}\log \ell}{Z}
    \int_{1/\ell}^{\infty} u^{-\alpha}-(1+u)^{-\alpha} du \\
  & + \frac{\ell^{1-\alpha}}{Z} \int_{1/\ell}^{\infty} (u^{-\alpha}-(1+u)^{-\alpha}) 
  \log (u^{-\alpha}-(1+u)^{-\alpha}) du
  ~.
\end{align*}
The first of the two integrals can be evaluated explicitly as:
\begin{align*}
\int_{1/\ell}^{\infty} u^{-\alpha}-(1+u)^{-\alpha} du =
-\frac{\ell^{\alpha-1}}{1-\alpha} + \frac{\ell^{\alpha-1}}{1-\alpha} (\ell+1)^{1-\alpha}
  ~.
\end{align*}
So, that we find the first term's asymptotic behavior to be:
\begin{align*}
-\alpha \frac{\ell^{1-\alpha}\log \ell}{Z} \int_{1/\ell}^{\infty}
u^{-\alpha}-(1+u)^{-\alpha} du \sim -\alpha \log \ell
  ~,
\end{align*}
plus corrections of $o(1)$. One of the more notable corrections of $o(1)$ is
proportional to $\frac{\log \ell}{Z}$, which is $o(1)$ for $\alpha<1$ and
otherwise has a nonzero limiting value when $\ell \rightarrow\infty$.  

Surprisingly, the latter of the two integrals limits to a finite value for $\alpha<1$:
\begin{widetext}
\begin{align*}
\lim_{\ell\rightarrow\infty} \frac{\ell^{1-\alpha}}{Z}
  \int_{1/\ell}^{\infty} (u^{-\alpha}-(1+u)^{-\alpha})
  & \log (u^{-\alpha}-(1+u)^{-\alpha}) du \\
  & = (1-\alpha) \int_{0}^{\infty} (u^{-\alpha}-(1+u)^{-\alpha})
  \log (u^{-\alpha}-(1+u)^{-\alpha}) du
  ~,
\end{align*}
where we used $\lim_{\ell\rightarrow\infty} \frac{\ell^{1-\alpha}}{Z} = 1-\alpha$ for
$\alpha<1$. As a result, we find that:
\begin{align}
\frac{1}{Z}\int_0^{\infty} (\Phi(t)-\Phi(t+\ell)) & \log
(\Phi(t)-\Phi(t+\ell))dt
  \nonumber \\
  & = \begin{cases}
  -\frac{1}{2}\log \ell & \alpha=1 \\
  -\alpha\log \ell
  + (1-\alpha) \int_{0}^{\infty} (u^{-\alpha}-(1+u)^{-\alpha})
  \log (u^{-\alpha}-(1+u)^{-\alpha}) du & 0<\alpha<1
  \end{cases}
  ~,
\label{eq:Qhard}
\end{align}
plus corrections of $o(1)$.
Altogether, combining Eqs.~(\ref{eq:Q0})-(\ref{eq:Q2}) and (\ref{eq:Qhard})
into Eq.~(\ref{eq:EERenCT}), we recover Eq.~(\ref{eq:EE_FRP}) of the main text.
\end{widetext}

As discussed there, we still must evaluate $\EE(\ell)$ at $\alpha=1$. We focus
again on asymptotic expansions in $\ell$ and drop corrections to expressions
that do not contribute to $\EE$. When $\alpha=1$:
\begin{align*}
Z(\ell) = 1 + \sum_{n=1}^\ell \frac{1}{n} = \log \ell
  ~,
\end{align*}
plus corrections of $O(1)$. Next, we evaluate:
\begin{align*}
-\sum_{n=0}^{\ell} w(n) \log w(n) & = \sum_{n=1}^\ell \frac{\log n}{n} \\
  & = \sum_{n=2}^\ell \frac{\log n}{n}
  ~.
\end{align*}
Since $\frac{\log n}{n}$ is a monotone decreasing function with $n$, we lower- and upper-bound this sum using integrals:
$\int_2^{\ell+1} \frac{\log n}{n} dn \leq \sum_{n=2}^\ell \frac{\log n}{n} \leq
\frac{\log 2}{2} + \int_2^\ell \frac{\log n}{n}dn$. These are easily evaluated,
giving:
\begin{align*}
-\sum_{n=0}^{\ell} w(n) \log w(n) &= -\frac{1}{2}\log^2 \ell
  ~,
\end{align*}
plus corrections of $O(1)$. For other sums, we need an expression for $F(n)$:
\begin{align*}
F(n) & = w(n)-w(n+1) \\
     & = \begin{cases}
	 0 & n=0 \\ \frac{1}{n(n+1)} & n\geq 1
	 \end{cases}
	 ~.
\end{align*}
Then, we evaluate:
\begin{align*}
\sum_{n=0}^{\ell} (n+1) F(n)\log F(n)
  & = -2 \sum_{n=1}^\ell \frac{\log n}{n}
  + \sum_{n=1}^\ell \frac{\log (1+\frac{1}{n})}{n} \\
  & = \log^2 \ell
  ~,
\end{align*}
plus corrections of $O(1)$, where we have noted that $\sum_{n=1}^{\infty}
\frac{\log (1+\frac{1}{n})}{n}$ converges since $\int_1^{\infty} \frac{\log
(1+\frac{1}{x})}{x} dx$ converges. The next term takes the form:
\begin{align*}
(\ell+1) \sum_{\ell+1}^{\infty} F(n) \log F(n)
  & = -(\ell+1) \sum_{\ell+1}^{\infty} \frac{\log (n(n+1))}{n(n+1)}
  ~.
\end{align*}
We can bound the sum using $\int_{\ell+1}^{\infty} \frac{\log (n(n+1))}{n(n+1)} dn
\leq \sum_{\ell+1}^{\infty} \frac{\log (n(n+1))}{n(n+1)} \leq \frac{\log
(\ell^2+\ell)}{\ell^2+\ell} + \int_{\ell+1}^{\infty} \frac{\log (n(n+1))}{n(n+1)} dn$. These
integrals are both easily evaluated, revealing an asymptotic form of:
\begin{align*}
(\ell+1) \sum_{\ell+1}^{\infty} F(n) \log F(n) & = -2\log \ell
  ~,
\end{align*}
plus corrections of $O(1)$.
Finally, to evaluate the last term in the sum, we note that:
\begin{align*}
w(n) - w(n+\ell+1) & = \frac{1}{n(1+\frac{n}{\ell+1})} \\
                & = \frac{1/\ell+1}{\frac{n}{\ell+1} (1+\frac{n}{\ell+1})}
  ~,
\end{align*}
when $n\geq 1$. We define $x_n = \frac{n}{\ell+1}$ with $dx_n = \frac{1}{\ell+1}$
and write:
\begin{align*}
w(n) - w(n+\ell+1) = \frac{dx_n}{x_n (1+x_n)}
  ~.
\end{align*}
Then:
\begin{align*}
& \sum_{n=0}^{\infty} (w(n) - w(n+\ell+1)) \log (w(n) - w(n+\ell+1)) \\
  & ~= (1-w(\ell+1)) \log (1-w(\ell+1)) \\
  & \quad + \log dx_n \sum_{n=1}^{\infty} \frac{dx_n}{x_n (1+x_n)}
  + \sum_{n=1}^{\infty} \frac{\log (x_n (1+x_n))}{x_n (1+x_n)} dx_n
  ~.
\end{align*}
The first term is $o(1)$, since $\lim_{\ell \rightarrow\infty} (1-w(\ell+1)) \log
(1-w(\ell+1)) = 0$. We can view the other two sums as Riemann sums for integrals
$\int_{1/\ell}^{\infty} \frac{dx}{x(1+x)}$ and $\int_{1/\ell}^{\infty} \frac{\log
(x(1+x))}{x(1+x)} dx$ respectively, giving:
\begin{align*}
\sum_{n=1}^{\infty} \frac{dx_n}{x_n (1+x_n)} & = \log \ell
  ~,
\end{align*}
plus corrections of $o(1)$ and:
\begin{align*}
\sum_{n=1}^{\infty} \frac{\log (x_n (1+x_n))}{x_n (1+x_n)} dx_n =
-\frac{1}{2}\log^2\ell
  ~.
\end{align*}
plus corrections of $o(1)$.
Altogether, substituting the above expressions into Eq.~(\ref{eq:EEL2}) yields:
\begin{align*}
\EE(\ell) & = \log \log \ell - 2
  ~,
\end{align*}
plus corrections of $o(1)$. The various divergences of order $\log \ell$ all
cancel one another, but the divergence of $\log\log \ell$ due to the $\log
\ell$ divergence in $Z(\ell)$ remains, just as for the continuous-time case.  When $F(n)$ is monotone decreasing at some finite $N$ sufficiently rapidly, manipulations similar to those above imply that divergence in $\ECT$ is a sufficient condition for divergence in $\EE$.

\bibliography{chaos}

\end{document}